\documentclass[prb, preprint]{revtex4}

\usepackage{graphicx}
\usepackage{units}

\begin{document}
\title{Measuring Phonon Dephasing with \\ Ultrafast Pulses}


\author{F. C. Waldermann} 
\author{Benjamin J. Sussman}
\email{ben.sussman@nrc.ca}
\altaffiliation[Also at ]{National Research Council of Canada, Ottawa, Ontario K1A 0R6, Canada}
\author{J. Nunn}
\author{V. O. Lorenz}
\author{K. C. Lee}
\author{K. Surmacz}
\author{K. H. Lee}
\author{D. Jaksch}
\author{I. A. Walmsley}
\affiliation{Clarendon Laboratory, University of Oxford, Parks Road, Oxford, OX1 3PU, UK}
\author{P. Spizziri}
\author{P. Olivero}
\author{S. Prawer}
\affiliation{Center for Quantum Computer Technology, School of Physics,
The University of Melbourne, Parkville, Victoria 3010, Australia}

\begin{abstract}
A technique to measure the decoherence time of optical phonons in a solid is presented. Phonons are excited with a pair of time delayed 80~fs, near infrared pulses via spontaneous, transient Raman scattering.  The fringe visibility of the resulting Stokes pulse pair, as a function of time delay, is used to measure the phonon dephasing time.  The method avoids the need to use either narrow band or few femtosecond pulses and is useful for low phonon excitations. The dephasing time of phonons created in bulk diamond is measured to be $\tau=$6.8~ps ($\Delta \nu=1.56$~cm$^{-1}$).
\end{abstract}

\maketitle
\section{Introduction}

Phonons are a fundamental excitation of solids that are responsible for numerous electric, thermal, and acoustic properties of matter.  The lifetime of optical phonons plays an important role in determining these physical properties and has been the subject of extensive study. A technique to measure phonon dephasing times is presented here that utilizes ultrafast infrared laser pulses:  Transient Coherent Ultrafast Phonon Spectroscopy (TCUPS) offers a number of  conveniences for measuring phonon dephasing.  TCUPS utilizes commercially available ultrafast pulses (80~fs) and hence does not require a narrow band or extremely short lasers to achieve high spectral or temporal resolution.  As well, TCUPS is suitable for measurements in the single phonon excitation regime.   The large sampling area and long sampling distance increase the generated Stokes power and avoid sample heating, which is a concern for low temperature studies.  Diamonds are well known for their extraordinary physical properties \cite{Field1979} and, as well,  offer interesting prospects for use in Quantum Information applications \cite{Wrachtrup2006,Childress2006, Neumann2008,Waldermann2007}.  As such, diamond has been selected here as the material for demonstration of TCUPS.

Two methods have previously been utilized to measure phonon lifetimes: high-resolution Raman spectroscopy and differential reflectivity measurements. The first is the traditional technique, where the optical phonon lifetime is obtained from high-resolution linewidth measurements of the first-order Raman peak, usually conducted using narrowband excitation lasers and high-resolution spectrometers \cite{lbpb2000}. The alternative technique, working in the time-domain, can directly show the temporal evolution of the surface vibrations of solids  \cite{ckk1990}. A femtosecond pump pulse is used to excite a phonon population. The reflectivity (or transmittivity) of a subsequent probe pulse displays a time dependence that follows the vibrational frequency and phonon population.  This method was used to study the phonon decay in various solids \cite{cbkmddi1990}, their symmetry modes \cite{lyylkl2003}, and their interaction with charge carriers \cite{hkcp2003} and with other phonons \cite{bdk2000}. In these experiments, impulsive stimulated Raman scattering has been established as the coherent phonon generation mechanism \cite{sfigwn1985,lfgwfum1995}. 

The time-domain experiments utilize the impulsive regime, \textit{i.e}.~laser pulse lengths much shorter than the phonon oscillation period (inverse phonon frequency). This requirement can be challenging for the application of the differential reflectivity technique to materials with high phonon energies, as laser systems with very short pulse lengths are required (\textit{e.g.}, for diamond, sub-10~fs pulses are required to resolve a phonon frequency of 40~THz).  TCUPS operates in the transient Raman scattering regime, \textit{i.e.}, pulse lengths much shorter than the phonon decoherence time, which is usually orders of magnitude slower than the phonon oscillation period (about 25~fs for diamond) \cite{ihk2007}.  Stimulated Raman scattering, which implies large phonon excitations,  is often employed in dephasing measurements in order to achieve good signal-to-noise ratios.   High phonon population numbers, often referred to as \emph{hot phonons}, can be subject to an increased decay rate, as previously observed \cite{ylokl2002} for {GaN}. By contrast, TCUPS investigates the properties of a coherent phonon excitation by direct analysis of the Stokes spectra generated in the Raman process. The use of single photon detectors extends the sensitivity of the experiment to low phonon populations, including the single phonon level. 

\section{Experiment}

\begin{figure}\centering
\includegraphics[width=\columnwidth]{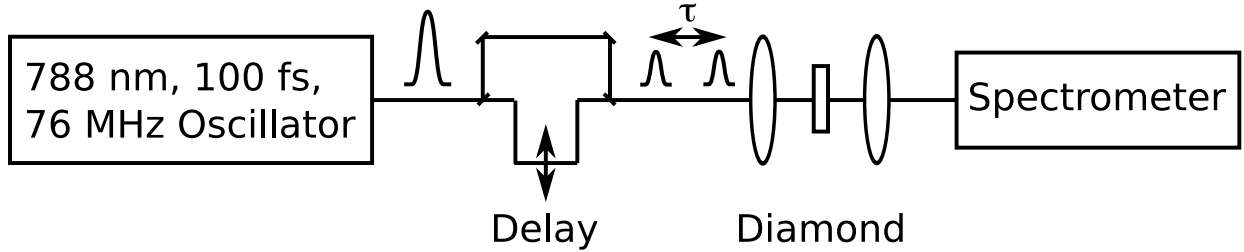}
\caption{Experimental setup.  An oscillator pulse is split into two time delayed pulses and focused through the diamond sample.  Not shown, a bandpass filter cleans the oscillator pulse before the diamond and a longpass filter rejects the pump and transmits the Stokes before the spectrometer.}
\label{fig:bulk_raman_experiment}
\end{figure}

\begin{figure}\centering
\includegraphics[width=\columnwidth]{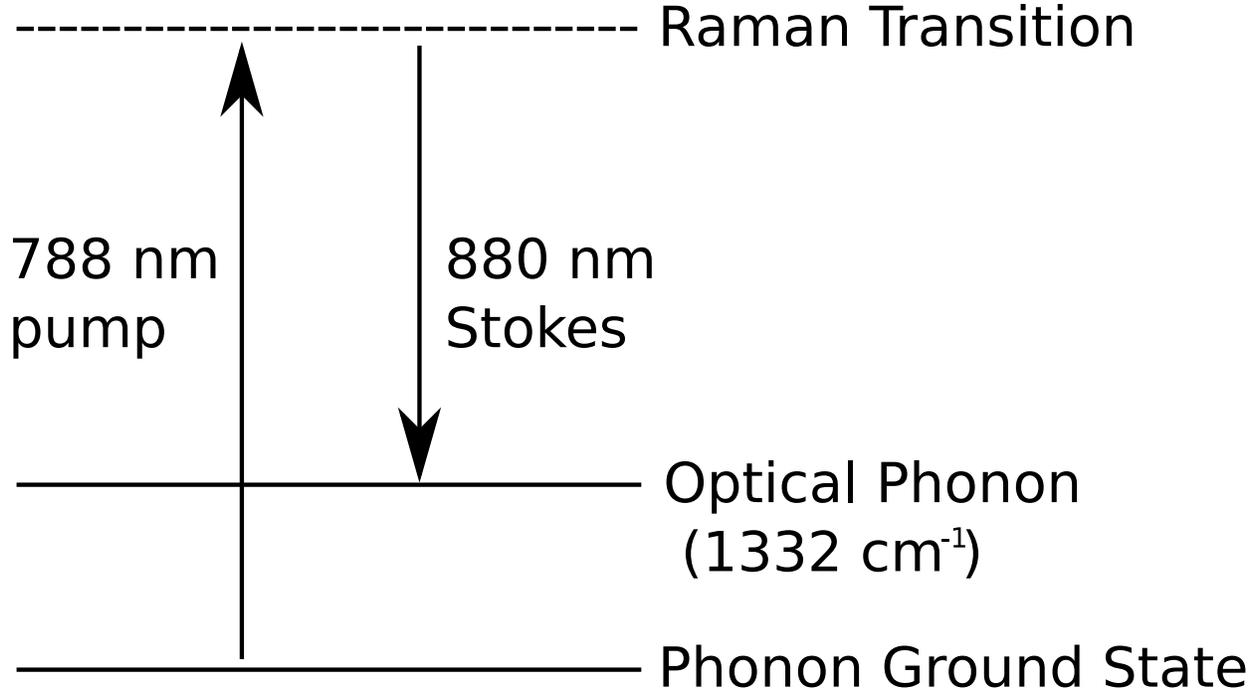}
\caption{Diamond energy level schematic.  Ground state phonons are excited with the incident 788~nm pump, via a Raman transition, to the optical phonon mode, emitting an 880~nm Stokes pulse}
\label{fig:diamondEnergyLevels}
\end{figure}

The diamond was classified as a type Ib high pressure high temperature (HPHT) sample with a nitrogen impurity concentration of less than 100~ppm. The Stokes shift of diamond \cite{zaitsev2001} is $\unit[1332]{cm}^{-1}$ and the Raman gain coefficient for diamond has been reported \cite{llk1971} as $g = \unit[7.4 \times 10^{-3}]{cm/MW}$ (corrected for $\lambda = \unit[800]{nm}$). With pump pulse energies ranging over $\unit[1.1 \ldots 380]{pJ}$, the collinear Stokes emission is calculated as $0.004 \ldots 1.3$ photons per pulse, in agreement with the count rates achieved experimentally. The Raman scatter is thus in the spontaneous regime, as verified by a linear pump power dependence ranging over three orders of magnitude (see inset of Fig.~\ref{fig:powerdep}). Therefore, the experiment is performed far below the hot phonon regime.

\begin{figure}[]
\centering
\includegraphics[width=\columnwidth]{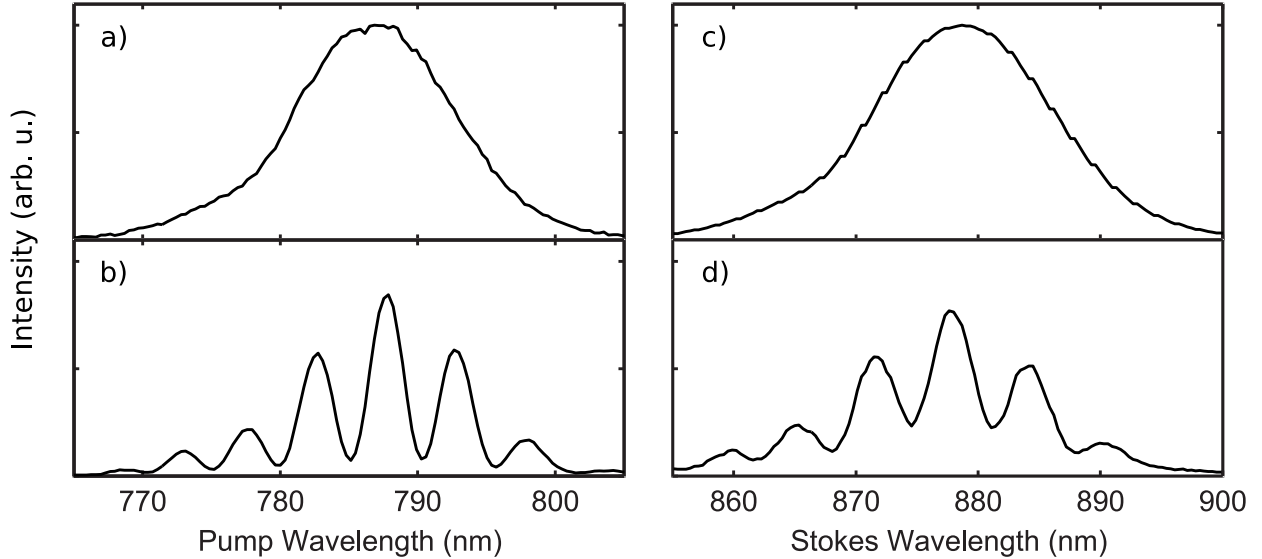}
\caption{Example spectral interference for a delay $\tau = 0.39$~{ps}. Spectra of the broadband excitation laser (left) and the Stokes signal of diamond (right). The single pulse data in (a) and (c) show the pump laser spectrum and the corresponding broadband Raman spectrum. Spectral interference fringes appear for coherent pulse pairs in (b) and (d). }
    \label{fig:two_pulse_spectra_principle}
\end{figure}

\begin{figure}[]
\centering
\includegraphics[width=\columnwidth]{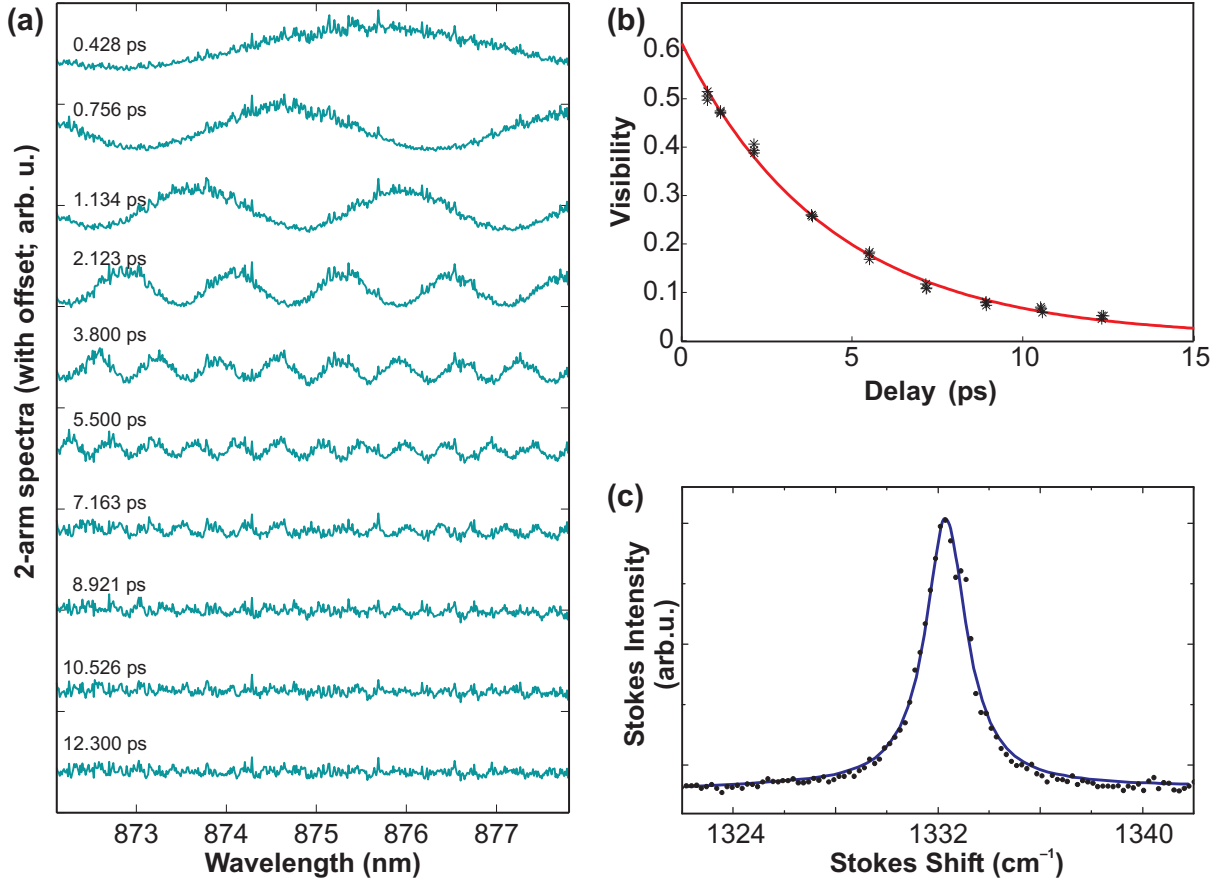}
\caption{Decoherence measurement. (a) Stokes spectra for two pump pulses with various delays $\tau$, recorded with a 1800 lines/mm grating. The decrease of spectral interference visibility of the Stokes signal is due to decoherence of the optical phonons created. The respective visibilities are plotted in (b), obtained by curve-fitting the spectra. Asterisks denote data points, the continuous line an exponential decay fit. Part (c) shows a high-resolution Raman spectrum for the same diamond (dots) with a Lorentzian fit (line). }
\label{fig:waterfall_etc}
\end{figure}

The experimental setup is depicted in fig.~\ref{fig:bulk_raman_experiment}. Phonons are excited, via a Raman transition, with a pair of time-delayed 80~fs, 788~nm pulses (fig.~\ref{fig:diamondEnergyLevels}) from a commercial Ti:Sapphire Oscillator (Coherent Mira).  The pulses are focussed into a $2 \times 2 \times 1$~mm diamond with faces polished along [100] plane (Sumitomo).  Stokes emission is detected collinearly. The pump laser is spectrally filtered using a bandpass filter to avoid extraneous light at the Stokes frequency, which might seed stimulated emission and decreases the signal-to-noise ratio when detecting single Stokes photons. The Stokes scatter is detected and spectrally analysed by means of a 30~cm spectrometer (Andor Shamrock 303i) and an electron multiplying charged coupled device (iXon DV887DCS-BV), which is capable of statistical single photon counting. The gratings were ruled at 150 lines/mm for data in figures~\ref{fig:two_pulse_spectra_principle}, and 1800 lines/mm for figures~\ref{fig:waterfall_etc} and~\ref{fig:powerdep}.

The spectral interference from the pump pair and Stokes pair is shown in fig.~\ref{fig:two_pulse_spectra_principle}.  The fringe spacing $\Delta \lambda$ is as expected for two time-delayed coherent pulses: $\Delta \lambda={\lambda^2}/{c \tau}$ (see also  (\ref{eq:IntensityFringes}), below).  For the excitation pair, $\lambda$ is the centre wavelength of the pump (fig.~\ref{fig:two_pulse_spectra_principle}b) and for the generated output Raman pair, $\lambda$ is the centre wavelength of the Stokes (fig.~\ref{fig:two_pulse_spectra_principle}d).  The fringe spacing of the Raman output corresponds to the Stokes peak wavelength, confirming that the process is a measure of the coherence of the Raman process.   Figure \ref{fig:waterfall_etc}a shows the fringe visibility reduction as a function of time delay.  The fringe visibility $V=\exp(-\Gamma|\tau|)$ is plotted in  fig.~\ref{fig:waterfall_etc}b.  The visibility has been renormalized using the laser visibility for each delay to account for beam walk-off and the spectrometer resolution which artificially reduces visibility, due to a sampling effect from the finite pixel size of the spectrometer CCD. 

\section{Theory}
The observed interference visibility can be considered from two perspectives.  In the first, the visibility decay arises due to fluctuations of the phase in the classical fields.  Each input laser pulse excites optical phonons, via a Raman transition (fig.~\ref{fig:diamondEnergyLevels}), which in turn causes the emission of two Stokes pulses.  That is, the Raman interaction maps the electric field of the two input pulses to two output Stokes shifted pulses
\begin{equation}
E_{Stokes}=E_1(t)+E_2(t).
\end{equation}
The phase of the first Stokes pulse $E_1$ is determined spontaneously, but the phase (and amplitude at stimulated intensities) of the second pulse $E_2$ is influenced by the coherence maintained in the phonon of the system following the first pulse, so that the output field may also be rewritten as
\begin{equation}
E_{Stokes}(t)=E_1(t)+e^{i \theta}E_1(t-\tau)
\end{equation} 
where $\tau$ is the time delay between the input pulses and $\theta$ is the spontaneously fluctuating phase difference between the pulses.  The spectral intensity of the Stokes pulse pair 
\begin{equation}
|E_{Stokes}(\omega)|^2=2|E_1(\omega)|^2 \left(1+\cos({\omega \tau +\theta})  \right )
\end{equation}
contains interference fringes whose position depends on the relative phase $\theta$.  Shot-to-shot, decoherence causes spontaneous fluctuations in $\theta$ and the fringe pattern loses visibility.  At longer delays $\tau$, the fluctuations increase, eventually reducing the visibility of any integrated fringe pattern to zero.  Assuming a Lorentzian lineshape with width $\Gamma$ for the distribution of the phase shift, the shot-to-shot averaged spectral intensity is broadened to
\begin{equation} \label{eq:IntensityFringes}
\left<|E_{Stokes}(\omega)|^2 \right>_{shots}=2|E_1(\omega)|^2 \left(1+e^{-\Gamma|\tau|}\cos({\omega \tau})  \right ).
\end{equation}
The phase fluctuations cause a reduction of the fringe visibility.

Alternatively, the fluctuating phase perspective can be connected with the second, quantum field perspective.  This formalism can also be made applicable in the stimulated regime.  The observed spectral intensity expectation value is proportional to the number of Stokes photons:
\begin{equation}
\left <|E_{Stokes}(\omega)|^2 \right>\propto \left <  A^{\dagger} A\right > 
\end{equation}
where the lowering operator $A(\omega)$ is a sum of the first $A_1$ and second $A_2$  pulse mode lowering operators:
\begin{equation}
\left<|E_{Stokes}(\omega)|^2 \right>\propto\left <  A_1^{\dagger} A_1\right > +\left <  A_2^{\dagger} A_2\right > +2 \Re \left <  A_1^{\dagger} A_2\right >.
\end{equation}
The final, correlated term (\textit{cf}. the decay term in (\ref{eq:IntensityFringes})) measures the phonon coherence that remains in the system between pulses.  During the evolution of the system, the starting time phonon mode $B^\dagger(0)$ is `mixed' into the Stokes photon modes $A_i$ due to application of the laser field.  The mixed-in term is then subsequently the source for spontaneous emission.  The source term is the same for both pulses, but during the period between pulses the coherence is reduced due to crystal anharmonicity and impurities.   For the correlation, the relevant terms to lowest perturbative order are (see appendix for the equations of motion)
\begin{equation}
A_1\approx A_1(0) -i g \tau_{pump} B^\dagger(0)
\end{equation}
\begin{equation}
A_2\approx A_1(0)e^{i \omega \tau} -i g\tau_{pump} B^\dagger(0)e^{-\Gamma \tau}
\end{equation}
from which the correlation term can be evaluated as 
\begin{equation}
  \left <  A_1^{\dagger} A_2\right > \approx g^2\tau_{pump}^2 \left <  B(0) B(0) ^\dagger \right > =   g^2 \tau_{pump}^2\left < N_B(0) +1 \right > e^{-\Gamma \tau} 
  \end{equation}
where $N_B(0)$ is the initial number of phonons, which in this case is the nearly zero thermal population.  This result links the phonon decoherence rate $\Gamma$ with the fluctuating phase perspective  linewidth $\Gamma$ from (\ref{eq:IntensityFringes}).
Therefore, measuring the reduction of the fringe visibility is a direct measure of the phonon dephasing time.

\section{Discussion}

The TCUPS measurement indicates a phonon dephasing time of $1/\Gamma=6.8\pm0.9$~ps or a linewidth of $\Delta \nu=1.56$~cm$^{-1}$.  The literature has reported a great deal of variation in linewidth measurements for diamond, varying from at least 1.1~cm$^{-1}$ to as high as $4.75$~cm$^{-1}$  \cite{lbpb2000,llk1971,Chen1995} .  
Here, the TCUPS (fig.~\ref{fig:waterfall_etc}b, $\Delta \nu=1.56$~cm$^{-1}$ ) and conventional Raman spectrum (fig.~\ref{fig:waterfall_etc}c, $\Delta \nu=1.95$~cm$^{-1}$) show comparatively good agreement.  The lifetime measured here is slightly shorter than the decay rate calculated theoretically by Debernardi \emph{et al.} \cite{dbm1995} for an ideal crystal ($\Delta \nu= \unit[1.01]{cm}^{-1}$ or $1/\Gamma=\unit[10.5]{ps}$), as the decay process is enhanced by lattice imperfections, vacancies, and the high concentration of substitutional nitrogen atoms, as is typical for this sort of diamond. The decay model considering acoustic phonon modes suggests that this deviation from the theoretical optimum is due to inhomogeneous broadening rather than additional pure dephasing. Future work will reveal whether ultra-pure diamond with very low crystal defect density can achieve a longer phonon lifetime. The creation of coherent phonons in diamond is heralded by the emitted Stokes photon, which could be employed for quantum optical experiments operating at room temperature like, \textit{e.g.}, schemes that transfer optical entanglement to matter\cite{maku2004,crfpek2005}.

\begin{figure}
\centering
\includegraphics[width=\columnwidth]{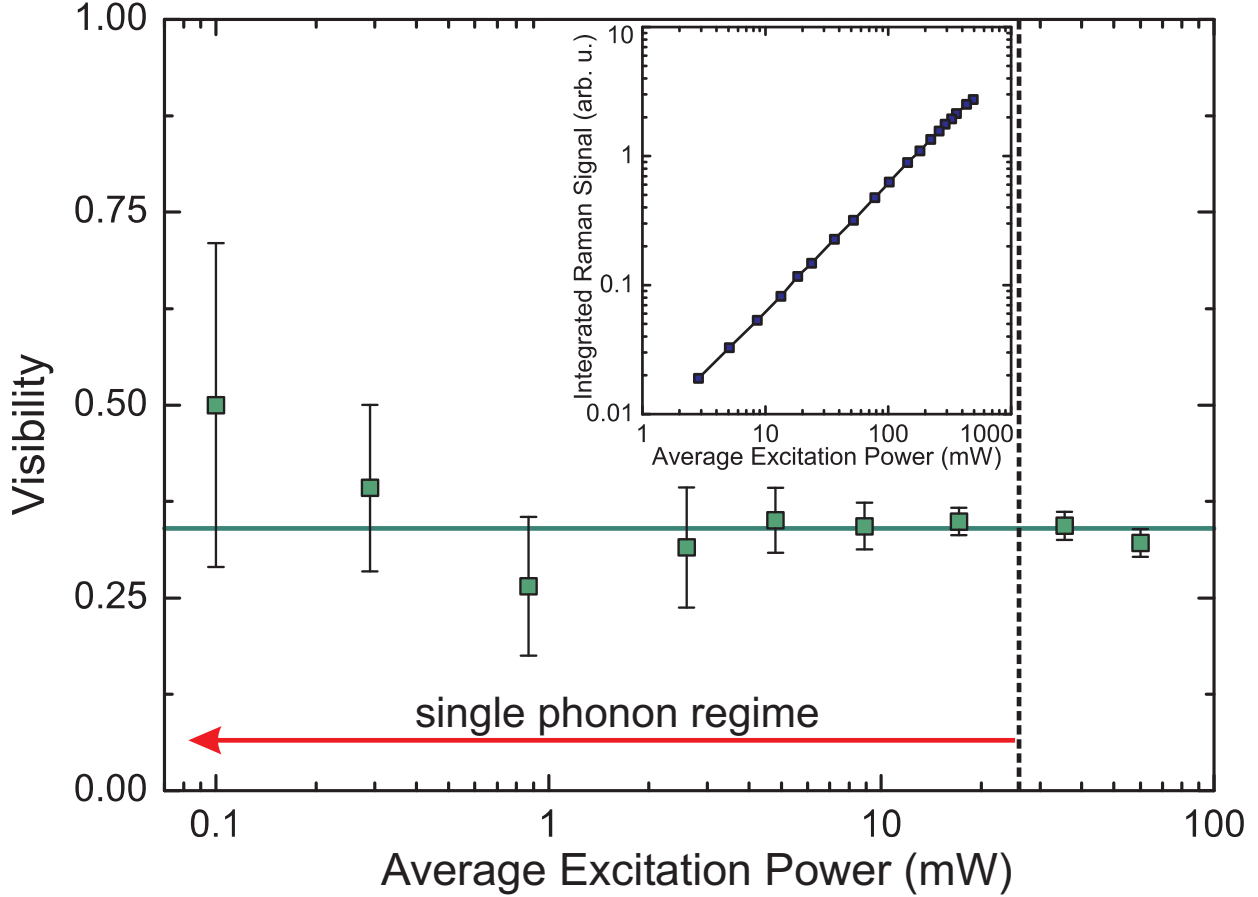}
\caption{Power dependence. The power dependence of the Stokes interference visibility (at constant delay $\tau = \unit[0.51]{ps}$, reduced due to a limited alignment) showing that the experiment can be carried out at arbitrarily low phonon excitation levels (the green horizontal line is plotted to guide the eyes). The inset shows the dependence of the Stokes pulse energy on the average pump power (single pump pulses).
The linear power dependence shows that the scattering is in the spontaneous Raman regime. The fraction of pump power converted into collinear Stokes light was measured to be less than $10^{-8}$.}
\label{fig:powerdep}
\end{figure}

The spectral interference pattern persists for low excitation levels, \textit{i.e.,}~a phonon excitation probability per mode of $p < 1$. (The case of $p \gg 1$ would correspond to a strongly stimulated regime, which has previously been studied in molecular hydrogen gas \cite{bswr1993}, although using spatial, not spectral interference.)  A constant visibility with excitation power can be seen at low excitation level (fig.~\ref{fig:powerdep}).  TCUPS can therefore be employed to measure the decoherence properties of single optical phonons, overcoming the need for large phonon populations for lifetime measurements of phonon decoherence.

The excitation probability per mode was much smaller than 1, ranging over $p \approx 10^{-7} \ldots 10^{-5}$ due to the large number of phonon modes in the Brillouin zone for which Stokes scatter is detected ($\sim 10^5$). This level is in fact smaller than the thermal population level of the optical phonons at room temperature, given by $p_\mathrm{thermal} = [\exp(E_\mathrm{vib} / k_B T) - 1]^{-1} \approx 0.0017$. However, a small thermal population of the optical phonon modes does not influence this measurement method, as only the phonons deliberately excited by the pump pulses lead to Stokes scatter, and only Stokes light and its interference are detected. As phonons are governed by bosonic statistics, any finite background excitation level does not inhibit a further excitation. The linewidth increase due to phonon-phonon interaction is negligible at ambient temperatures in diamond due to the low population level \cite{dbm1995}. An increase in the Raman linewidth of diamond due to temperature has been reported \cite{lbpb2000} to begin at around $T \approx \unit[300]{K}$. At $T \approx \unit[800]{K}$, it is more than twice the zero temperature linewidth. At room temperature, the phonon decay is only marginally  enhanced by an acoustical phonon population. This insensitivity to a thermal background is in contrast to the differential reflectivity method, where thermal phonons lead to additional noise as both thermal and coherent phonons  lead to a change of the material reflectivity. 

TCUPS is a convenient approach to determining the quantum coherence properties of optical phonons in Raman active solids. The  measurement technique relies solely on spontaneous Raman scattering and is therefore useful down to the single phonon levels. In particular, TCUPS enables the measurement of the decoherence time of phonons, which is of paramount importance in many Quantum Information Processing schemes. Spectral interference of the Stokes light from pump pulse pairs is used to measure the Raman linewidth of the material, while maintaining a coherent excitation due to ultrafast excitation. The phonon lifetime of diamond was measured as $\unit[6.8]{ps}$. This lifetime corresponds to a phonon $Q$-factor of $Q = \nu /\Gamma \sim 270$. Although the short lifetime of the excitation makes it unsuitable for long-distance quantum repeaters, such a  high $Q$ and the low thermal population at room temperature make it feasible for proof-of-principle demonstrations of typical quantum optics schemes, such as collective-excitation entanglement in the solid state. 

Acknowledgements. This work was supported by the {QIPIRC} and {EPSRC} (grant number GR/S\-82176\-/\-01), EU RTN project EMALI, and Toshiba Research Europe.

\section*{Appendix: Phonon-photon equations of motion}

Consider an incident pump laser that Raman scatters off a phonon field of the diamond to produce an output Stokes field. The equations of motion for Stokes field $A(t)$ and the phonon field $B(t)$ are linked by the pump coupling $g$ via \cite{Raymer1990}:
\begin{equation}
\dot{A}(t)=-ig B^\dagger(t)
\end{equation}
and
\begin{equation}
\dot{B}(t)=-igA^\dagger(t)-\Gamma B(t) + F^\dagger(t).
\end{equation}
The dephasing rate $\Gamma$ is due to crystal anharmonicity and impurities.  The Langevin operator $F$ has been added to maintain the normalization of $B$ in the presence of decay, allowing the phonon to decohere, but keeping the operator norm, via the commutation relation $[B,B^\dagger]=1$, constant.
The formal solutions are:
\begin{equation}
B=B(0)e^{-\Gamma t} +\int_0^t  e^{-\Gamma (t-t^\prime)}\left [ -i g A^\dagger(t^\prime)+F^\dagger(t^\prime)\right ] dt^\prime
\end{equation}
and
\begin{equation}
A=A(0)-i\int_0^t  g  B^\dagger(t^\prime)dt^\prime.
\end{equation}
For brevity, the time argument has been dropped from the solutions.  In the weak ($g \tau_{pump} \ll 1$) and transient ($\Gamma \tau_{pump}\ll 1$)  pump pulse limit, the incident laser leaves the phonon operator approximately in the vacuum state $B(0)$ and the phonon operator solution at lowest order is
\begin{equation}
B\approx B(0)e^{-\Gamma t} +\int_0^t  e^{-\Gamma (t-t^\prime)} F^\dagger(t^\prime)  dt^\prime.
\end{equation}
The Stokes field to first order is then
\begin{equation}
A \approx A(0) -i g \tau_{pump}B^\dagger(0)e^{-\Gamma t} -i \int_0^t \int_0^{t^\prime} g  e^{-\Gamma (\tau-t^{\prime \prime})} F(t^{\prime \prime}) dt^{\prime \prime} dt^\prime 
\end{equation}
where the coupling $g$ in the second term has been taken as a constant step for the duration of the pump. The initial Stokes operator $A(0)$ annihilates the vacuum,  but the solution for $A$ mixes in a component of the phonon raising operator $B^\dagger(0)$, which acts as a source for the spontaneous Raman scattering.

The decoherence rate $\Gamma$ represents the dephasing of the phonon raising operator $B^\dagger$.  The phonon number  $N_B=B^\dagger B$ therefore decays at a rate $2\Gamma$.  The corresponding spectral frequency linewidth  is $\Delta \nu={\Gamma}/{\pi}$.

\bibliographystyle{unsrt} 
\bibliography{fw_articles}

\end{document}